
\nopagenumbers
\magnification=\magstep1

\def\DD{{\rm D}}
\def\haf{{1\over2}}
\def\dd{{\rm d}}
\def\ee{{\rm e}}
\def\ssc{\scriptscriptstyle}

\def\dd{\hbox{d}}
\def\tr{\hbox{tr}}
\def\prl#1{$^{#1}$}
\font\tif=cmr10 scaled\magstep3

\def\al{\alpha}
\baselineskip=22truept
\rightline{iassns-hep-93-30}
\vskip -4truept
\rightline{hep-th/9305115}
\vskip -4truept
\rightline{5/93}
\vfil
\centerline{\tif Topological closed-string interpretation}
\centerline{\tif of}
\centerline{\tif Chern-Simons theory }
\vfil
\centerline{Vipul Periwal\footnote{${}^\dagger$}{vipul@guinness.ias.edu}}
\bigskip
\centerline{The Institute for Advanced Study}
\centerline{Princeton, New Jersey 08540-4920}
\vfil
\par\noindent
The exact free energy of SU($N$) Chern-Simons theory at level $k$
is expanded in powers of $(N+k)^{-2}.$  This expansion
keeps rank-level duality manifest,  and simplifies
as $k$ becomes large, keeping $N$ fixed (or vice versa)---this is
the weak-coupling (strong-coupling) limit.
With the standard normalization, the free energy on the three-sphere
in this limit is shown to be the generating function of the Euler
characteristics of the moduli spaces of surfaces of genus $g,$
providing a string interpretation for the perturbative expansion.
A similar expansion is found for the three-torus, with
differences that shed light on contributions from different
spacetime topologies in string theory.
\vfil\eject
\baselineskip=14truept
\def\th{1}
\def\prec{2}
\def\big{10}\def\w{6}\def\g{3}\def\gt{5}\def\joe{4}
\def\clz{11}\def\kac{9}\def\dual{7}\def\penn{8}
\def\vafa{13}
\def\qcd{$\hbox{QCD}_2$}
\def\sun{SU($N$)}
The perturbative expansion of any quantum field theory(qft) with fields
transforming in the adjoint representation of \sun\ is
a topological expansion\prl\th\ in surfaces, with $N^{-2}$ playing the r\^ole
of a handle-counting parameter\prl\prec.  For $N$ large, one hopes that the
dynamics of the qft is approximated by the sum (albeit largely
intractable) of all planar
diagrams.  The topological classification of diagrams has nothing
{\it a priori} to do with approximating the dynamics with a theory
of strings evolving in spacetime.

Gross\prl\g\ (see also ref.'s \joe, \gt) has shown recently that
the large $N$ expansion does actually provide a way of
associating a theory of strings in \qcd.  Maps of
two-dimensional string worldsheets into two-dimensional
spacetimes are necessarily somewhat constricted.
What one would like is a qft with fields
transforming in the adjoint representation in $d>2,$ which is
at the same time exactly solvable.  One could then, in principle,
attempt to associate a theory of strings with such a qft by exhibiting
a `sum over connected surfaces'
interpretation for the free energy of the qft.
There is no guaranty that such an association will exist.

Chern-Simons theory in three dimensions
is precisely such a rara avis among qfts.
It is described by a functional integral
$$Z[M]\ \equiv \ \int \DD A \exp\left(ik I_{\ssc CS}\right),$$
with $I_{\ssc CS} \equiv {1\over 4\pi}
\int_M \dd^3x \tr \left[A\dd A + {2\over 3}
A^3\right].$  Here $M$ is a closed oriented 3-fold, and the gauge group
is assumed simply-connected so the principal $G$-bundle on which $A$ is a
connection can be trivialized.  The trace is normalized so $k$ is an
integer.  For $G=$\sun, the case considered in detail in this Letter, the
trace is taken in the defining representation.  In
the standard large $N$ limit, $k\propto N.$
The functional integral will not be used in the following.
Instead, I shall exploit Witten's\prl\w\
relation of Chern-Simons theory to knot theory and conformal field theory.

I shall show in this Letter that the exact free energy of
\sun\ Chern-Simons theory at level
$k$ has an expansion that admits a string interpretation.  This
expansion is obtained in two steps, the first an
expansion in $(k+N)^{-2},$ and the second a `double-scaling'
limit\prl\big\nobreak.  The expansion in $(k+N)$ is natural from the point of
view of rank-level duality\prl\dual.  When $M$ is the three-sphere, $S^3,$ the
scaled free energy turns out to be the generating function of
the Euler characteristics of the moduli spaces of surfaces with $g$
handles\prl\penn.  I shall also compute the expansion when $M$ is the
three-torus, $T^3.$  This can be interpreted as a sum over surfaces with
one boundary (or puncture).  An important comment: the normalization
of the partition functions used in this Letter is that used in
Chern-Simons theory, and may be inappropriate for string
identifications.  For example, comparison with Casson's invariant
suggests that the free energy on $S^3$ should be chosen to vanish.

\def\al{\alpha}
\def\be{\beta}

To start, Witten\prl\w\ showed that
$$Z[S^3]\ =\ S_{0,0},$$
with the normalization $Z[S^2\times S^1] = 1.$
Here $S_{\al}^{\be}$ is the modular transformation matrix representing
the action of the modular group SL$(2,{\bf Z})$ on the characters of
the \sun\ Kac-Moody algebra at level $k:$
$$\chi_{\ssc\al}(-{1/\tau}) = \sum_\be S_{\al}^{\be} \chi_{\ssc\be}(\tau).$$
The characters of the WZW model are known, so $S$ can be determined
without resorting to computations with the functional integral
definition of the Chern-Simons theory.  Thus for \sun\prl\kac, we have
$$Z[S^3,N,k] = (k+N)^{-N/2} \sqrt{k+N\over N}
\prod_{j=1}^{N-1}\left\{2\sin\left({j\pi \over N+k}\right)\right\}^{N-j}.$$
The expansion and scaling limit undertaken below lead to
a result that can be deduced simply by taking a
large $k$ limit of $Z[S^3,N,k].$  The longer
route makes contact with the idea of double-scaling\prl\big\ and
keeps rank-level duality manifest in the scaling limit.
Define $M\equiv k+N,$ and
$x\equiv N/M.$  The rank-level duality\prl{\dual} under
an interchange of $N$ and $k,$
or equivalently $x\leftrightarrow 1-x,$ is easily checked:
$${Z[S^3,N,k]\over Z[S^3,k,N]} = \sqrt{1-x\over x }.$$
The normalization of the partition functions could be changed
to give exact duality; however, the normalization we use is natural from
the point of view of Kac-Moody algebras.
The planar term in the large $N$ expansion with $x$ fixed is
$$ \ln Z \asymp N^2 \int_0^1 (1-y) \ln2\sin\left(\pi yx\right),$$
obtained by Camperi, Levstein and Zemba\prl\clz.
\def\ww{12}

Recall
$\sin(\pi z) = \pi z \prod_{n=1}^\infty \left(1-{z^2/ n^2}\right).$
It follows that
$$Z[S^3,N,k] = M^{-((Mx)^2-1)/2} (2\pi)^{Mx(Mx-1)/2} G(Mx+1)
\exp(-F_{\ssc 0}),$$
where $G$ is Barnes' $G$-function\prl\ww, and
$$\eqalign{
F_{\ssc 0} &\equiv \sum_{n=1}^\infty\sum_{j=1}^{Mx-1}\sum_{m=1}^\infty
{1\over m}(Mx-j) \left({j\over nM}\right)^{2m}\cr
&= \sum_{m=1}^\infty {1\over m} {\zeta(2m)\over M^{2m}}\sum_{j=1}^{Mx-1}
(Mx-j) j^{2m}.\cr}$$
The sum over $j$ can be written in terms of the Bernoulli
polynomials\prl\ww,
$$\phi_{k}(z) \equiv z^k - {k\over 2} z^{k-1} + C^{k}_2 B_1 z^{k-2}
- C^{k}_{4} B_{\ssc 2} z^{k-4} + \dots \hbox{up to $z$ or $z^2$},$$
which satisfy
$$\sum_{j=1}^{L-1} j^{m-1} = {1\over m} \phi_m(L).$$
After some rearrangement, we arrive at
$$\eqalign{ F_{\ssc 0} =
M^2 &\sum_{m=1}^\infty {\zeta(2m)\over m} {x^{2m+2}\over
(2m+1)(2m+2)}\cr
- &\sum_{k=1}^\infty M^{2-2k} (-)^k B_k {{2k-1}\over
2k!} {\dd^{2k-2}\over{\dd x^{2k-2}}} \ln \Big({{\sin\pi x}\over{\pi
x}}\Big).\cr}$$
The asymptotic behaviour of Barnes' $G$-function is
$$\ln G(z+1) \asymp {z\over 2}\ln 2\pi - {3z^2\over 4} +{z^2\over2}\ln z
-{{B_{\ssc 1}}\over 2} \ln z +\sum_{r=2}^\infty (-)^{r-1} {{B_{\ssc r}}\over
2r(2r-2)} {z^{2-2r}} .$$
We can now put all the terms together, noting that the $\ln\pi x$
terms from $F_0$ cancel against terms coming from the $G$-function.
This renders the rank-level duality manifest in the higher genus
contributions.  The end result, if we fix $N\equiv Mx,$ and take
$M\uparrow\infty,$ is
$$\eqalign{F[S^3,N] = {1\over 2} \ln x
&+ N^2\Big[{3\over 4} - {1\over 2}\ln 2\pi x\Big]
+{B_1\over 2} \ln N \cr
&+\sum_{k=2}^\infty N^{2-2k} (-)^k {B_k\over2k(2k-2)}.\cr}$$
The (virtual) Euler characteristic of surfaces with $g$ handles\prl\penn\ is
$$\chi_g \ = \ (-)^g{{ B_g}\over 2g(2g-2)}.$$
Thus, even including the scaling violations evident in the $N^2 \ln x$
term\prl\vafa,
the free energy we have computed is precisely the generating function
of Euler characteristics.  Thus we have found a string
interpretation of the free energy of Chern-Simons theory, since the
Euler class is a natural measure on moduli space, and one way of
characterizing string theories is to give measures on moduli spaces for
all genus.
\def\haf{{1\over 2}}

I turn now to the three-torus, $T^3.$  Witten\prl\w\ showed that
$Z[T^3,N,k]$  counts the number of integrable irreducible highest-weight
representations of the affine Kac-Moody algebra at level $k.$
These are in one-to-one correspondence with Young tableaux
with at most $N-1$ rows and at most $k$ columns.
We thus find
$$Z[T^3,N,k] = {1\over kB(k,N)},$$
where $B$ is the Euler Beta-function.  Observe that
$Z[T^3,N,k]/Z[T^3,k,N] =x/1-x,$ similar to the analogous ratio for
$S^3.$ The $B$-function
can be written as
$$B\left(M(1-x),Mx\right) \ =\ x^{Mx-\haf}(1-x)^{M(1-x)-\haf}
\sqrt{2\pi\over M} \exp C(M,x),$$
with
$$ C(M,x) \ = \ 2\int_0^\infty \ {{\dd t}\over{\ee^{2\pi t}-1}}
\arctan\left[{{(tM^{-1})^3 +tM^{-1}(1-x+x^2)}\over{x(1-x)}}\right].$$
In the limit $Mx\equiv N$ fixed, $M\rightarrow\infty,$ using
$B_k = 4k \int_0^\infty \dd t\ t^{2k-1} \left(\ee^{2\pi t} -1\right)^{-1},$
and $\arctan(z) \approx z - z^3/3 +z^5/5 - \dots \ (z\approx 0),$
we find
$$F[T^3,N,k] \asymp -{1\over 2}\ln\left({{ N}\over2\pi }\right)
- N (1-\ln x) +\sum_{k=1}^{\infty} (-)^{k} N^{1-2k}
{B_k\over 2k(2k-1)}.$$
Thus the expansion should be interpreted as a sum over surfaces with
one puncture or hole.  However, the Euler characteristic of surfaces
with $m$ punctures and $g$ handles is
$$\chi_{g,m} = (-)^m {\Gamma(2g-2+m)\over\Gamma(m+1)\Gamma(2g-2)}\chi_g,$$
thus the coefficient of $N^{1-2k}$ is not as we would have expected for
surfaces with a simple puncture.  Also observe the unexplained
$\ln N$ term which is the only term that does not come from a surface
with just one puncture.

Why is the appearance of a boundary
natural? Every 3-fold can be obtained from
$S^3$ by cutting out a tubular neighbourhood of
links embedded in $S^3,$ and then gluing the tubular neighbourhood back
after acting on its boundary by a
diffeomorphism---this is called surgery.  $T^3$ can be obtained from
$S^3$ by surgery on Borromean rings embedded in $S^3.$
We would ideally wish to
associate terms in the perturbative expansion on $T^3$  with surfaces whose
boundaries are the reglued Borromean rings in $S^3.$
However, the Borromean rings are linked and there is no simple
way (without self-intersection) that one can make surfaces
with Borromean ring boundaries.

\def\lisa{14}
\def\wtwo{15}
\def\wthre{16}
It is of great interest to understand the
Chern-Simons string theory for different 3-folds
since it may teach us how to obtain a
formulation of string theory independent of spacetime topology.
It should be possible to explicitly
compute the free energy on many simple 3-folds
following the general results found by Witten\prl\w.  For example,
if one defines a function
$$\Xi(s) \equiv \ \sum \ {S_{0,\al}}^{-s},$$
where the sum runs over all integrable irreducible
highest weight representations,
then $\Xi(2g-2)$ is the partition function on $\Sigma_g\times S^1.$
Jeffrey's\prl\lisa\
Poisson resummation method should be useful in this analysis.
For some idea of the computations involved, see ref.~\wtwo.
Witten gave an open string interpretation to
Chern-Simons theory\prl\wthre.  The appearance of the Euler
characteristics in the present work seems unrelated to
the explicit geometry of his
construction, but this merits further investigation.

\bigskip
I am grateful to S. Shatashvili and E. Witten for encouragement and
helpful conversations.  I thank P. Aspinwall for help in computing
the partition function on $T^3,$ and L. Jeffrey for references.
This work was supported by D.O.E. grant DE-FG02-90ER40542.

\medskip

\centerline{References}
\medskip
\item{\th .} G. 't Hooft, {\sl Nucl. Phys.} {\bf B72} (1974) 461
\item{\prec .} To be precise, the action defining the qft should be of
the form ${N\over g^2}\int \tr \left[\phi K\phi +V(\phi)\right],$
with $K$ the inverse propagator, and $g$ is independent of $N.$
\item{\g .} D.J. Gross, Princeton preprint PUPT 1356 (1992)
\item{\joe .} J.A. Minahan, Virginia preprint UVA-HET-92-10 (1992)
\item{\gt .} D.J. Gross and W. Taylor IV, Princeton preprint 1376
(1993)
\item{\w .} E. Witten,
{\sl Comm. Math. Phys.} {\bf 121} (1989) 351
\item{\dual .} S. Naculich, H. Riggs and H. Schnitzer,
{\sl Phys.  Lett.} {\bf 246B} (1990) 417;
M. Camperi, F. Levstein and G. Zemba,
{\sl Phys.  Lett.} {\bf 247B} (1990) 549; earlier work found
rank-level duality in RSOS models, A. Kuniba and T. Nakanishi,
`Level-rank duality in fusion RSOS models', in Proc. Int. Coll. on
Modern quantum field theory, (TIFR, Bombay, 1991), and in WZW
models, S. Naculich and H. Schnitzer, {\sl Phys. Lett.} {\bf 244B}
(1990) 235, {\sl Nucl. Phys.} {\bf B347} (1990) 687
\item{\penn .} J. Harer and D. Zagier, {\sl Invent. Math.} {\bf 185}
(1986) 457;
R.C. Penner, {\sl Bull. Amer. Math. Soc. (N.S.)} {\bf 15} (1986) 73,
{`The moduli space of punctured surfaces'}, preprint
\item{\kac .} V.G. Kac and M. Wakimoto, {\sl Adv. Math.} {\bf 70} (1988)
156
\item{\big .} E. Br\'ezin and V. Kazakov, {\sl Phys. Lett.} {\bf 236B}
(1990) 144; M. Douglas and S. Shenker, {\sl Nucl. Phys.} {\bf
B335} (1990) 635; D. Gross and A. Migdal, {\sl Phys. Rev. Lett.} {\bf
64} (1990) 127
\item{\clz .} M. Camperi, F. Levstein and G. Zemba,
{\sl Phys.  Lett.} {\bf 247B} (1990) 549
\item{\ww .} E.T. Whittaker and G.N. Watson, {\it A course of modern
analysis}, Cambridge Univ. Press, Cambridge (1980)
\item{\vafa .} J. Distler and C. Vafa, in {\it Random surfaces and
quantum gravity},  eds. O. Alvarez et al. (Plenum Press, New York,
1991)
\item{\lisa .} L. Jeffrey,
{\sl Comm. Math. Phys.} {\bf 147} (1992) 563
\item{\wtwo .} E. Witten,
{\sl Comm. Math. Phys.} {\bf 141} (1991) 153
\item{\wthre .} E. Witten, IAS preprint IASSNS-HEP-92/45 (1992)
\end